\documentclass{book}
\def\shorttitle{Two Greatest Mysteries}
\usepackage{narosa}
\begin{document}
\title{FOREWORD:\\
Probing the Two Greatest Mysteries of the Universe}
\author{D.P. Roy}
\affil{Homi Bhabha Centre for Science Education, Tata Institute of Fundamantal Research,
Mumbai - 400088, India \\
Instituto de Fisica Corpuscular, CSIC-U de Valencia, Correos E-46071 Valencia, Spain}

\beginabstract

The neutrinoless double beta decay and the direct dark matter
detection experiments probe the origins of the two geatest mysteries
of the universe, i.e. the baryon asymmetry and the invisible or dark
matter. The underlying theoretical ideas are briefly discussed along
with the experimental prospects.

\endabstract

One of the main achievements of modern physics is the unification of the microcosm with the macrocosm. It follows from the basic principles of uncertainty and mass energy equivalence that when we probe deep into the subatomic space we come across states of very high energy and mass. These are the particles that abounded in the early history of the universe, when its ambient temperature was as high as their mass. We have now a good understanding of these subatomic particles up to a mass range of 100 GeV in terms of the so called standard model. This has helped us to trace back the history of the universe to within a few picoseconds of its creation, when the ambient temperature was about a 100 GeV. Moreover we are now probing physics beyond the standard model mass scale of 100 GeV and looking into the very early history of the universe, when it was much less than a picosecond old. In the process we hope to understand the two greatest mysteries of the Universe, originating from that very early history -- i.e. (1) the baryon asymmetry and (2) the dark (or invisible) matter.

\section*{Baryon Asymmetry of the Universe (BAU) :}
The visible universe is made of baryon (proton, neutron) and lepton (electron) with hardly any evidence of their antiparticles. This is a mystery since the standard model (SM) treats them symmetrically. As per the SM the basic constituents of matter are a dozen of fermions along with their antiparticles; i.e. the three pairs of leptons
$$ \pmatrix{ \nu_e \cr e} \pmatrix{ \nu_\mu \cr \mu} \pmatrix{\nu_\tau \cr \tau} $$
and three pairs of quarks, which constitute the baryons. They have three basic interactions -- strong, electromagnetic and weak. These are all mediated by gauge bosons with couplings proportional to the corresponding gauge charges -- i.e. colour charge, electric charge and weak isospin. Parity is broken maximally by the weak, but conserved by strong and electromagnetic interactions. Accordingly only the left-handed leptons and quarks occur as weak isodoublets, carrying isospin  1/2, as represented by the above pairs. On the other hand the right-handed charged leptons and quarks occur as isosinglets, carrying no isospin. It is exactly the opposite for the antiparticles, since all these gauge interactions preserve CP symmetry. Since the neutrino has no strong or electromagnetic interaction, there is no right-handed neutrino in the SM.

The mass term corresponds to a coupling of left- and right-handed fermions, which does not conserve isospin. Thus in the SM the matter fermions get their masses from spontaneous breaking of the weak gauge symmetry by the Higgs mechanism, which also gives mass to the weak gauge bosons. Therefore all these masses are limited by the symmetry breaking scale of about 100 GeV. But the neutrino has no mass in the SM since there is no right-handed neutrino. However, we know from the neutrino oscillation experiments now that the neutrinos have tiny but nonzero mass, much less than an eV, i.e. more than a billion times smaller than the other fermion masses. 

A simple and elegant explanation of such tiny neutrino masses comes from the See-Saw model. It goes a step beyond the SM and assumes that the neutrino has a right-handed singlet component like the other fermions. Unlike the other fermions, however, the singlet neutrino is seen to have a unique property that it carries no gauge charge. Thus it is identical to its antiparticle, except for their opposite chiralities. This is called a Majorana fermion. It has a Majorana mass from the coupling of the left-handed neutrino with its right-handed antiparticle. Of course it breaks lepton number conservation, i.e. $\Delta L = 2$. But since the lepton number is not a gauge charge of the SM, this does not break any symmetry of the SM. So the Majorana mass M can be much larger than the SM symmetry breaking scale, i.e. $M >> 10^2$ GeV. Moreover, like the other fermions  one also gets a normal (Dirac) mass m from the coupling of the right-handed singlet with the left-handed doublet  neutrinos, which is limited by this symmetry breaking scale. Thus in the basis of left and right chirality one gets a 2x2 neutrino mass-matrix
$$
\pmatrix{ 0 & m \cr m & M}$$                                 
with $M >> m$. On diagonalization one gets the two mass eigenvalues as 
$M$ and $m^2/M$. In contrast the other fermions have no Majorana mass, so that on diagonalization both the right and left handed fermions have the same Dirac mass m. So it is the more than a billion times larger Majorana mass of the right-handed singlet neutrino, that is responsible for pushing down the left-handed doublet neutrino mass more than a billion times lower than the other fermion masses. This is why it is called the See-Saw model. Thus the measurement of the ultralight neutrino masses in the neutrino oscillation experiments provides us indirect evidence for the ultraheavy Majorana masses, which can not be probed directly in any foreseeable future. Finally the lepton number violation, $\Delta L = 2$, associated with the ultraheavy Majorana mass can be responsible for generating a lepton asymmetry in the very early history of the universe, when its age was many billion times smaller than a pecosecond. Then the quantum anomaly associated with the SM could convert a part of this lepton asymmetry into a baryon asymmetry at a slightly later stage, when its age was a few picoseconds. This is in fact the leading model for explaining the BAU, which is evidently a very important matter as it underlies our very existence in the universe. But we have still a long way to go in order to prove (or disprove) this model. 

\begin{figure}[!htb]
\vskip 0in
\vspace*{7mm}
\hspace*{12mm}
Normal hierarchy: \hspace*{2.8cm} Inverted hierarchy:
\vspace*{3mm}
\vskip .2in
\epsfxsize=5.5cm\epsfbox{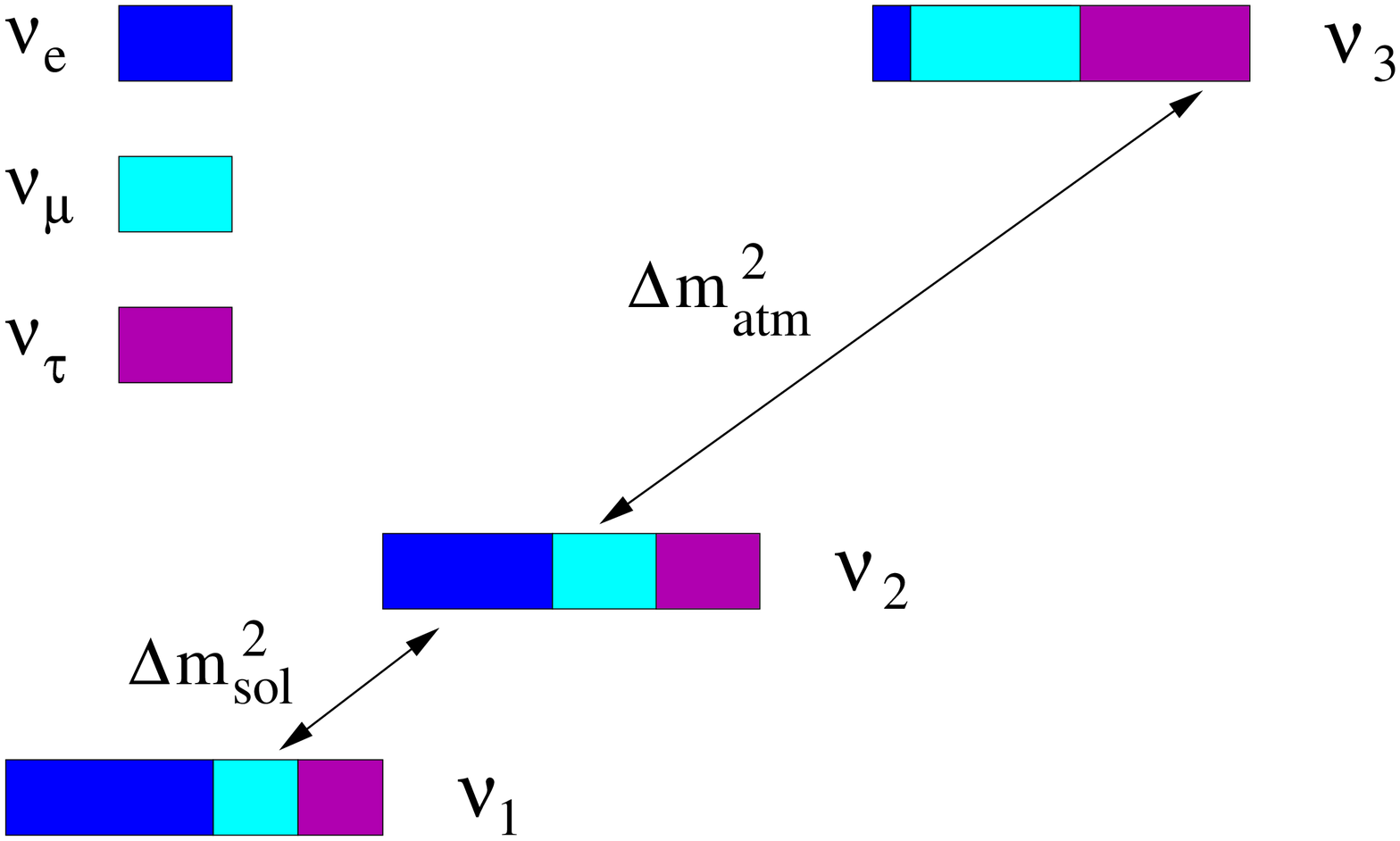}
\hspace{4mm}
\epsfxsize=5.55cm\epsfbox{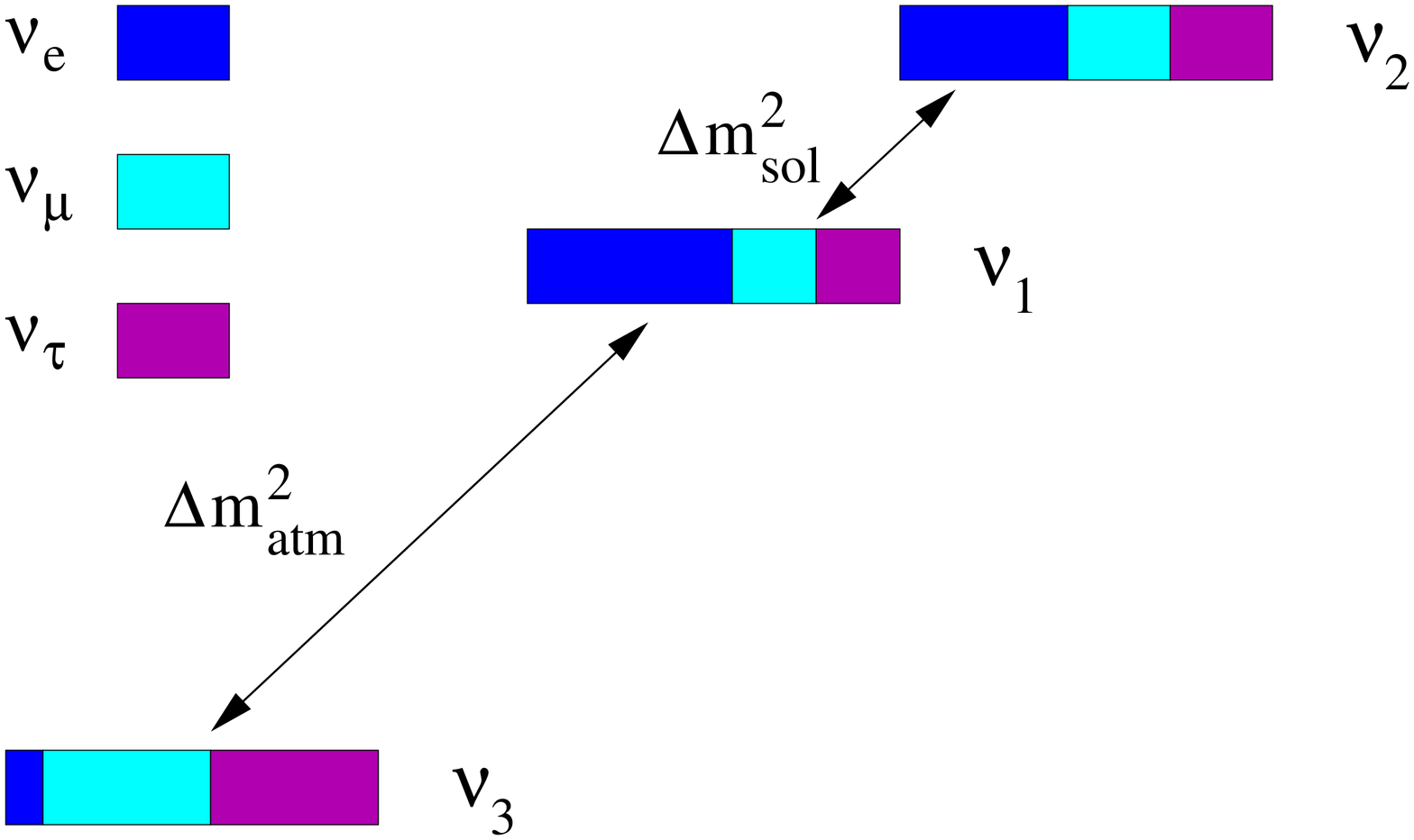}
\caption{\label{fg1}  A schematic diagram of nutrino masses and mixings}
\vskip 0.2in
\end{figure}

Figure \ref{fg1} illustrates what we know about the masses and mixings of the three light neutrinos from atmospheric and solar neutrino oscillation data along with those from accelerator and reactor neutrino experiments (E. K. Akhmedov, hep-ph/0610064).

There are two mass-square gaps $\Delta m^2$ 
between the three neutrino masses. Thanks to the solar matter effect, we know both the magnitude and the sign of the smaller gap $m_{sol}
\simeq 8 \times 10^{-5}~eV^2$, while we know only the magnitude of the larger gap $m_{atm} \simeq 2.5 \times 10^{-3}~eV^2$. We also know two of the three mixing angles, which are both large; but we have only a limit on the third one representing $\nu_e$ content  of the 3rd mass eigenstate, i.e. $|U_{e3}^2|
< 0.05$. The sign of the $m_{atm}^2$ leads to a two-fold ambiguity called normal and inverted hierarchies, as illustrated above. Secondly we do not know the absolute mass scale -- the lowest mass could be larger than the gaps implying relatively large and degenerate neutrino masses. Third and most important, we do not know for sure the Majorana nature of these masses, since the neutrino oscillation involves no L violation. So we can not rule out the possibility that the neutrino has no Majorana mass, but its Dirac mass is much smaller than those of other fermions for some as yet unknown reason. 

The neutrinoless double beta deacay (NDBD) experiment holds the key to all these three issues. Many experiments have observed double beta decay between  neighbouring even-even nuclei,
$$
{}^A_Z X \to {}^A_{Z+2} X + 2 e^- + 2 \bar \nu_e
$$
If neutrinos have Majorana mass then the associated lepton number violating mechanism can lead to pair-annihilation of the two antineutrinos, leading to the NDBD
$$
{}^A_Z X \to {}^A_{Z+2} X + 2 e^- ,
$$
which will show up as a peak in the total kinetic energy of the electron pair. The rate of the NDBD is proportional to the square of the Majorana mass
$$
| <m_{ee}>| = |\sum_i m_i U_{ei}^2 |,
$$
where the constant of proportionality comes from the nuclear matrix elements. They can be estimated from processes like the first one, although there is still considerable uncertainty in their estimates. The NDBD experiment of the Moscow-Hedelberg collaboration has already limited the above Majorana mass to $<m_{ee}> < 0.35$ eV. Assuming Majorana nature of neutrino mass this gives arguably the best limit on the mass of each neutrino in the degenerate mass scenario. In the inverted hierarchy scenario the two $\nu_e$ rich mass eigenstates of the above figure have atmospheric scale mass of 0.05 eV. The next generation of NDBD experiments propose to extend the probe of neutrino Majorana mass down to this level, i.e. they can cover the inverted hierarchy along with the degenerate mass scenarios. If they find no NDBD signal then the final step will be to extend the probe down to the solar neutrino mass scale of about 0.01 eV, corresponding to the mass of the $\nu_e$ rich states in the normal hierarchy scenario. At this level the NDBD experiment will be able to either confirm or disprove the Majorana nature of neutrino mass. This last step may be a long way off; but there is no other way for this lofty goal.

\section*{Dark Matter of the Universe (DM) : }

We know for a long time now that the ordinary baryonic matter constitutes only a small fraction of the matter content of the universe, while most of it comes from an invisible matter of unknown origin, which is called the dark matter (DM) for historical reason. The oldest and the strongest evidence for DM comes from the rotational velocity curves of stars or hydrogen clouds, lying outside the visible disc of galaxies, as shown in the following figure.. Balancing the centrifugal acceleration with the gravitational attraction of the galaxy gives
$$
{v^2 \over R} = {G_N M(R) \over R^2} \Longrightarrow v^2 = 
{G_N M(R) \over R}
$$

\begin{figure}[!htb]
\vskip 0in
\epsfxsize=110mm
\centerline{\epsfbox{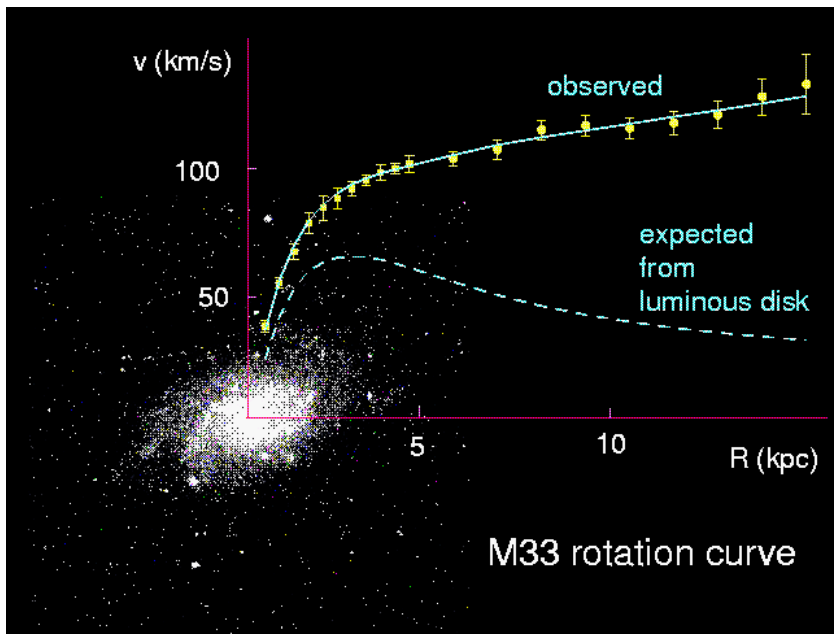}}
\vskip 0in
\caption{Rotation velocity curve of the nearby dwarf spiral galaxy M33, superimposed on its optical image ( E. Corbelli and P.Salucci, astro-ph/9909252)}
\end{figure}

If there were no galactic matter outside the visible disk then the rotation velocity curve would have gone down like $1/\sqrt{R}$. Instead it continues to rise way beyond the visible disk, suggesting that there is a lot of invisible matter in and around the galaxy. Similar results have been obtained from about a thousand galaxies, including our own, as well as cluster of galaxies. 

A natural candidate for this invisible matter comes from supersymmetry, which is invoked to stabilise the symmetry breaking scale of the SM against divergent quantum corrections. It is the lightest suppersymmetric particle (LSP), which is a colourless and chargeless object having only weak interaction like the neutrino, but with a mass of $ 10^{2-3}$ GeV. It is also called  weakly interacting massive particle (WIMP). This particle can be looked for at the forthcoming large hadron collider at CERN, Geneva, at least over a part of the supersymmetric parameter space. But whether such particles are seen at CERN or not, their connection with the invisible or dark matter of the universe can only be established through the direct dark matter detection (DDMD) experiments. It is based on measuring the nuclear recoil energy in the elastic scattering of the DM on a heavy nucleus. Since the recoil energy comes from the kinetic energy of the DM, which is only $10^{1-2}$ KeV, one needs to have a very low energy threshold.

\section*{Concluding Remarks :} 
As we have seen above the NDBD and the DDMD experiments
probe the origins of the two greatest mysteries of nature, i.e. the baryon asymmetry and the dark matter of the universe. They are both challenging experiments. The reaction rates are very low, which calls for large target mass. Besides one has to contend with high cosmic ray and radioactive backgrounds, which call for deep underground location and selection of extremely radiopure materials and environment. Not withstanding these challenges, there are about a dozen of DDMD experiments around the world and about twice as many NDBD experiments, including the complete and the upcoming ones. This shows the degree of global importance and priority enjoyed by these two experiments. It may be noted here that these challenges being common to both NDBD and DDMD experiments, very often they face it together by sharing the same underground laboratory and the same radiopure environment. Moreover some of the NDBD experiments, which are based on bolometric technique, have been able to achieve sufficiently low energy threshold and carry on DDMD simultaneously. I have had the opportunity to see two such experiments. One was the abovementioned Moscow-Heidelberg experiment in the Gran-Sasso road tunnel in Italy, which has simultaneously looked for both NDBD and DDMD. The second was the dark matter research centre in a Hydroelectric power tunnel in South Korea, which is engaged in both these searches.

Finally I compliment the organisers of this dedicated workshop on NDBD and DDMD to have brought together theorists and experimentalists from India and abroad to take stock of the present status and future prospects of these two very important experiments. And I conclude with the hope that this will help to advance the nascent effort of the Indian physics community to initiate these experimental activities in this country.

\end{document}